\documentstyle [11pt] {article}

\def\L{{\cal L}}

\begin{document}
\baselineskip=24pt

\title{{\normalsize {\bf SPACETIMES ADMITTING A 3-PARAMETER SIMILARITY GROUP}}}

\author{ {\normalsize J. Carot, L. Mas and A.M. Sintes }}
\date{} 
\maketitle
\centerline{ Dept. F\'{\i}sica}
\centerline{ Universitat Illes Balears}
\centerline{ E-07071 Palma de Mallorca. SPAIN}

\vskip4cm

\centerline{ \Large{ \bf Abstract}}
\vskip1cm

Spacetimes admitting a similarity group are considered. Amongst them,  special
attention is given to the 3-parameter ones. A classification of  such spacetimes
is  given  based  on the  Bianchi type of the similarity group $H_3$, and the
general form of the metric is provided in each case assuming the orbits are
non-null.

\setcounter{page}{0}
\newpage

\section{Introduction}

Spacetimes admitting (local) groups of (local) isometries \cite{Kramer} have
been widely studied, especially in those cases where the dimension of the Lie
algebra of Killing vectors fields (KV) is high or where there is  a 
non-trivial isotropy subgroup. On the other hand, Hall and Steele \cite{Hall90}
have investigated those spacetimes admitting an r-dimensional Lie algebra of
homothetic vector fields (HVF) (which gives rise to an r-parameter group of
similarities); solving the problem completely in those cases where $r\geq 6$
and also giving some general results when $r \leq 5$. The purpose of this
paper is, up to a certain extent, to complement the study carried out in the 
above reference, especially in the case $r=3$, which could be of interest in
Cosmology (see, for instance \cite{Hewitt88,Hewitt90,Hewitt91} and references
cited therein).

Throughout this paper $(M,g)$ will denote a spacetime: $M$ then  being  a
(smooth) Hausdorff,  simply connected 4-dimensional manifold, and $g$ a
(smooth) Lorentz metric with signature (-+++). A semicolon will denote a
covariant derivative with respect to the metric connection associated with $g$,
and a comma  will denote a partial derivative as usual. A global vector field
$X$ on $M$ is called homothetic if either of the two equivalent conditions 
\begin{eqnarray}
\L_Xg_{ab}\equiv X_{a;b}+X_{b;a}=2 \lambda g_{ab} \nonumber\\
 X_{a;b}= \lambda g_{ab}+F_{ab} \quad (F_{ab}=-F_{ba})
\label{1}
\end{eqnarray}
 holds on a local chart, where $\lambda$ is a constant on $M$, $F$ is the
homothetic bivector, and $\L$ denotes the Lie derivative operator. If
$\lambda\ne 0$, $X$ is called proper homothetic and it can always be scaled so
as to have  $\lambda=1$, if $\lambda = 0$ then $X$ is a KV on $M$. For a
geometrical interpretation of (\ref{1}) we refer the reader to
\cite{Hall88,Hall90b}.

A necessary condition that $X$ be homothetic is
\begin{equation}
{X^a}_{;bc}={R^a}_{bcd}X^d
\label{2}
\end{equation}
where ${R^a}_{bcd}$ are the components of the Riemann tensor in a coordinate
chart; thus, an HVF is a particular case of affine collineation \cite{HallCos88}
and therefore it will satisfy
\begin{equation}
\L_X {R^a}_{bcd}=\L_X R_{ab}=\L_X {C^a}_{bcd}=0
\label{3}
\end{equation}
where $R_{ab}$ ($\equiv {R^c}_{acb}$) and ${C^a}_{bcd}$ stand, respectively, for
the components of the Ricci and the Conformal Weyl tensor. Also, from the
Einstein's Field equations (EFE), it follows
\begin{equation}
\L_X T_{ab}=0
\label{4}
\end{equation}
where $T_{ab}$ is the energy momentum tensor representing the material content
of the spacetime ($M,g$). It can easily be shown that whenever a proper HVF
exists in a Lie algebra of HVF's ${\cal H}_r$, this necessarily  contains  an
$(r-1)$-dimensional Lie subalgebra of KV ${\cal G}_{r-1}$; therefore one can
always choose a basis for  ${\cal H}_r$ in such a way that it contains at most
one proper HVF, the $r-1$ remaining ones thus being KV's. If these vector fields
in the basis of  ${\cal H}_r$ are all complete vector fields, then  ${\cal
H}_r$ gives rise in a well known way to a Lie group of homotheties; otherwise,
it gives rise to a local group of local homothetic transformations of $M$ and,
although the usual concepts of isotropy and orbits still hold, a little more
care is required. We shall not go into the details here,  but refer the reader
to \cite{Hall90} for further information on this particular issue.
It is also immediate to see from (\ref{1}) that the Lie bracket of a proper HVF
and a KV is a KV. Further information on the isotropy structure
as well as on the fixed point structure of $H_r$ can be found in
\cite{Hall90,Hall88} (see also \cite{Beem}).

In this paper we shall be concerned with spacetimes admitting a 3-parameter
group of homotheties acting on non-null orbits, providing a classification of
all possible Lie algebra structures (in terms of the Bianchi type of ${\cal
H}_3$), and giving in each case the form of the metric as well as that of the
proper HVF and the two KV's, in terms of local coordinates. We shall also
present a few selected examples of spacetimes, satisfying the above
properties, which can represent perfect fluid cosmological models, as well as
vacuum solutions.

The paper is organized as follows: Section 2 contains a brief summary of
results on groups of homotheties and spacetimes admitting them. In section 3
we present the Bianchi classification of Lie algebras ${\cal H}_3$ along with
some remarks on the topology of the orbits of the corresponding Lie
subalgebras ${\cal G}_2$. The general form of the metric is provided in each
case. Finally, section 4, contains a few selected examples of perfect fluid
spacetimes admitting a maximal $H_3$; some of them we believe are new and,
whenever this is possible, we relate our results to those already existing in
the literature.

\section{Preliminary Results}

In this section we provide (without proof) some general results regarding
spacetimes admitting Lie groups of HVF. In what is to follow we shall be
assuming that a proper HVF exists; $r$ will then denote the dimension of the Lie
algebra of HVF ${\cal H}_r$ and ${\cal G}_{r-1}$ will be its associated Killing
subalgebra. Most of the following results, together with their proofs, can be
found in ref. \cite{Hall90}. \begin{enumerate}
\item The orbits of  ${\cal H}_r$ and ${\cal G}_{r-1}$ can only coincide if
they are 4-dimensional or 3-dimensional and null. Thus, the case in which we
will be interested mainly ($r=3$, non-null orbits) corresponds to a transitive
action of the homothety group (and therefore, the dimensions of the orbits of
${\cal H}_3$ and ${\cal G}_2$ will be 3 and 2 respectively).
\item If $r=11$ then $M$ is flat. The cases $r=10$, $9$ are impossible, as it
follows from consideration of the dimension of ${\cal G}_{r-1}$. $r=8$
corresponds to $M$ being a conformally flat, homogeneous generalized plane
wave \cite{Kramer}. The case $r=7$ implies that $M$ is a type $N$, homogeneous
or a conformally flat non-homogeneous generalized plane wave or one of the
special Robertson-Walker spacetimes (or their equivalent, with Segre type
$\{(1,11)1\}$ for the Ricci tensor). $r=6$ implies that $M$  is a type $N$,
non-homogeneous, generalized plane wave. In the $r=5$ case, the associated
${\cal G}_4$ subalgebra has necessarily 3-dimensional non-null orbits, the
Petrov type being $D$, $N$ or $O$ for timelike Killing orbits, and $D$ or $O$
for spacelike ones. \item If $r=4$ and a multiply transitive action is assumed,
then ${\cal H}_4$ and ${\cal G}_3$ have respectively 3-dimensional and
2-dimensional orbits. \item Spacetimes admitting a 4-parameter group of
homotheties acting transitively on $M$ were studied by Rosquist and Jantzen
\cite{Rosq} (see also \cite{McIntosh} where a thorough study of vacuum Bianchi
$I$ solutions admitting a proper HVF is carried out).
 \item The case $r=3$ has an associated Killing subalgebra ${\cal G}_2$ and the
respective dimensions of their orbits are 3 and 2 (see remark (1) above). In
this case one can classify the Lie algebras  ${\cal H}_3$ according to their
Bianchi type (see for example \cite{Petrov}); the only possible types  being
those corresponding to soluble groups ($I$ to $VII$ in the previous reference),
as it follows from the fact that  ${\cal H}_3$ must contain a 2-dimensional
subalgebra  ${\cal G}_2$, which in all cases but one, turns out to be abelian.
In this case (abelian ${\cal G}_2$), there are only two different topologies
possible for the (non-null) orbits $V_2$; namely: $V_2$ diffeomorphic to ${\cal
R}^2$, and $V_2$ diffeomorphic to  ${\cal S}^1\times {\cal R}$; and it follows
that in the latter case \cite{Mars} the only Bianchi type possible for  ${\cal
H}_3$ is $I$; as for the case $V_2\cong {\cal R}^2$, all seven
types can, in principle, occur.
 \end{enumerate}

\section{Bianchi types of ${\cal H}_3$}

The purpose of this section is to analyze the possible Bianchi types of ${\cal
H}_3$, giving in each case the coordinate forms of the proper HVF and the
metric tensor. We shall restrict ourselves to the case of non-null orbits, and
furthermore we shall assume (as is customary) that the Killing orbits $V_2$
admit orthogonal 2-surfaces. We shall denote the KV's spanning ${\cal G}_2$ 
by $\xi$ and $\eta$, and the proper HVF in the basis of  ${\cal H}_3$ as $X$;
also we shall treat separately the case where  ${\cal G}_2$ is abelian from
that where  ${\cal G}_2$ is non-abelian.

\subsection{Case  ${\cal G}_2$ abelian}

Under the above assumptions, the possible Bianchi types of  ${\cal H}_3$
containing an abelian  ${\cal G}_2$ are \cite{Petrov}:\hfill\break
\begin{eqnarray}
(I)&\quad &[\xi,\eta]=[\xi,X]=[\eta,X]=0 \nonumber \\
(II)&\quad &[\xi,\eta]=[\xi,X]=0 \quad [\eta,X]=\xi  \nonumber \\
(III)&\quad &[\xi,\eta]=0\quad [\xi,X]=\xi \quad [\eta,X]=0  \nonumber \\
(IV)&\quad &[\xi,\eta]=0\quad [\xi,X]=\xi \quad [\eta,X]=\xi+\eta  \nonumber \\
(V)&\quad &[\xi,\eta]=0\quad [\xi,X]=\xi \quad [\eta,X]=\eta  \nonumber \\
(VI)&\quad &[\xi,\eta]=0\quad [\xi,X]=\xi \quad [\eta,X]=q\eta  \nonumber \\
(VII)&\quad &[\xi,\eta]=0\quad [\xi,X]=\eta \quad [\eta,X]=-\xi+q\eta \quad 
(q^2<4) \nonumber 
\end{eqnarray}

Assume now that the Killing orbits $V_2$ are spacelike and diffeomorphic to
${\cal R}^2$; since $\xi$ and $\eta$ commute, we locally have
\begin{equation}
\xi={\partial \over \partial x} \qquad \eta={\partial \over \partial y}
\label{5}
\end{equation}
taking now two more coordinates, $t$ and $z$, it follows from the assumption
that the Killing orbits $V_2$ admit orthogonal surfaces, that the line 
element associated to the metric $g$ can be written as
\begin{equation}
ds^2=\Psi^{-2}\{ -dt^2+dz^2+s^2dy^2+b^2(Pdy+dx)^2\} 
\label{6}
\end{equation}
where $\Psi$, $s$, $b$ and $P$ are all functions of $t$ and $z$ alone, their
functional dependence on these coordinates to be determined (to some extent) by
the  HVF $X$ in each case ($I-VII$).

For timelike Killing orbits, also diffeomorphic to ${\cal R}^2$, one (locally)
has: 
\begin{equation}
\xi={\partial \over \partial t} \qquad \eta={\partial \over \partial z}
\label{7}
\end{equation}
and the metric would then read:
\begin{equation}
ds^2=\Psi^{-2}\{ dx^2+dy^2+s^2dz^2-b^2(dt+Pdz)^2\} 
\label{8}
\end{equation}
where $\Psi$, $s$, $b$ and $P$ are now functions of $x$ and $y$, coordinates
on the surfaces orthogonal to the Killing orbits.

The case  $V_2\cong {\cal S}^1 \times{\cal R}$, either spacelike (cylindrical
symmetry) or timelike (stationary axially symmetric metric) is easily
obtained from (\ref{5}) and (\ref{6}) (respectively (\ref{7}) and (\ref{8}))
by simply changing $y$ (respectively $z$) to  $\varphi$, angular coordinate
(and then one has to impose the regularity condition on the axis,
ref.\cite{Kramer} p.192; in order to ensure that $\varphi$ has the standard
periodicity $2 \pi$). It is precisely the fact that the axis of rotation is an
(invariant) submanifold of the spacetime manifold, what implies that -if no
other CKV exists on $M$- then ${\cal H}_3$ must be abelian (Bianchi type $I$)
\cite{Mars}.

In what is to follow, and for the sake of simplicity, we shall assume that the
Killing orbits $V_2$ are spacelike and diffeomorphic to ${\cal R}^2$, so that
the forms (\ref{5}) and (\ref{6}) will hold for the KV's and the line element
respectively. The case of timelike Killing orbits can be formally obtained by
changing $(x,y)$ into $(t,z)$ in the expression (\ref{5}) for the KV's $\xi$
and $\eta$; and
\begin{equation}
t\longrightarrow ix\qquad y\longrightarrow z \qquad P\longrightarrow iP
\label{9}
\end{equation}
in (\ref{6}), to get the line element in this case.

Now, assume $X$ is a proper-HVF satisfying (\ref{1}) (with $\lambda=1$); in a
coordinate chart it will have an expression of the form
\begin{equation}
X=X^a(x^b)\partial_a
\label{10}
\end{equation}
Specializing now the equation (\ref{1}) to the HVF (\ref{10}) and the metric
(\ref{6}) one sees that, assuming non-null homothetic orbits $V_3$, it is always
possible to perform a coordinate change in the 2-spaces orthogonal to the
Killing orbits; such that preserves the form of the metric and brings the HVF
(\ref{10}) to one of the two following forms: \begin{equation}
X=\partial_t + X^x(x,y)\partial_x + X^y(x,y)\partial_y 
\label{11}
\end{equation}
or
\begin{equation}
X=\partial_z + X^x(x,y)\partial_x + X^y(x,y)\partial_y 
\label{12}
\end{equation}
Assuming the form (\ref{11}) for the proper HVF $X$ (i.e.: 3-dimensional
timelike homothetic  orbits), equation (\ref{1}), along with each particular
Lie algebra structure, yields for every different Bianchi type ($I$) to ($VII$)
the following forms for $X$ and the functions $\Psi$, $s$, $b$ and $P$
appearing in  (\ref{6}),

\begin{eqnarray}
(I) &\,& X=\partial_t \qquad\Psi=e^{-t}f(z)\quad s=\hat s(z) \quad b=\hat
b(z) \quad  P=\hat p(z) \qquad \label{13}\\
(II) &\, & X=\partial_t+y\partial_x \qquad\Psi=e^{-t}f(z)\quad s=\hat
s(z) \quad b=\hat b(z) \quad  P=\hat p(z)-t  \\
(III) &\, & X=\partial_t+x\partial_x \qquad\Psi=e^{-t}f(z)\quad s=\hat
s(z) \quad b=e^{-t}\hat b(z) \quad  P=e^t\hat p(z)  \\
(IV) &\, & X=\partial_t+(x+y)\partial_x+y\partial_y
\qquad\Psi=e^{-t}f(z)\\
 &\, &  s=e^{-t}\hat s(z) \quad b=e^{-t}\hat
b(z) \quad  P=\hat p(z)-t  \nonumber\\
 (V) &\, & X=\partial_t+x\partial_x+y\partial_y
\qquad\Psi=e^{-t}f(z)\label{17}\\
&\, & s=e^{-t}\hat s(z) \quad b=e^{-t}\hat b(z) \quad 
P=\hat p(z)  \nonumber\\
 (VI) &\, & X=\partial_t+x\partial_x+qy\partial_y\quad
(q\not=0,1) \qquad\Psi=e^{-t}f(z) \\
&\, & s=e^{-qt}\hat s(z) \quad b=e^{-t}\hat b(z) \quad 
P=e^{(1-q)t}\hat p(z)  \nonumber\\
(VII) &\,& X=\partial_t -y\partial_x+(x+qy)\partial_y\quad (q^2<4)
\qquad\Psi=e^{-t}f(z) \label{19}
\end{eqnarray}
$$ s={e^{-{q\over2}t}{\sqrt{4-q^2}\over
2}a(z) \over (\sqrt{a(z)^2+c(z)^2+g(z)^2} +c(z)\cos(\sqrt{4-q^2}t)
+g(z)\sin(\sqrt{4-q^2}t))^{{1\over2}}}$$
$$ b=e^{-{q\over2}t}(\sqrt{a(z)^2+c(z)^2+g(z)^2}
+c(z)\cos(\sqrt{4-q^2}t) +g(z)\sin(\sqrt{4-q^2}t))^{{1\over2}} $$
$$ P={q\over 2}+{\sqrt{4-q^2}\over
2}{-g(z)\cos(\sqrt{4-q^2}t)+ c(z)\sin(\sqrt{4-q^2}t) \over
\sqrt{a(z)^2+c(z)^2+g(z)^2} +c(z)\cos(\sqrt{4-q^2}t)
+g(z)\sin(\sqrt{4-q^2}t)}$$

If $P=$constant two mutually orthogonal KV's exist in ${\cal G}_2$ and one can
always, by means of a linear change of coordinates in the Killing orbits 
$V_2$, bring $P$ to zero. It is worth noticing that this is not  possible for
families ($II$) and ($IV$). Note that in case $VII$,  $P=$constant implies the
existence of a third KV tangent to the Killing orbits $V_2$, therefore these
orbits are of  constant curvature and the Lie algebra of HVF is 4-dimensional.

The other ansatz for the proper HVF $X$, (\ref{12}) corresponds to the
homothetic orbits being spacelike, and would yield similar results to those
above but with the role of the coordinates $t$ and $z$ reversed.

As for the case  $V_2\cong {\cal S}^1 \times{\cal R}$, and from the previous
remarks on this issue, it follows that the metric  would be that given in
(\ref{6})  (or (\ref{8})) exchanging $y$ (or $z$) for $\varphi$, and with
$\Psi$, $s$, $b$ and $P$ being those given in (\ref{13}) (or their equivalent
under the substitution (\ref{9}) in the case of timelike Killing orbits).

\subsection{Case ${\cal G}_2$ non-abelian}
The Lie algebra structure in this case is 
\begin{equation}
[\xi,\eta]=\xi \quad [\xi,X]=[\eta,X]=0
\label{20}
\end{equation}
Assuming that the theorem of Bilyalov \cite{Bilyalov} and Defrise-Carter
\cite{Defrise} holds (for a precise statement of the conditions under which
this happens, see \cite{Hall90b}), there exists a (smooth) function
$\sigma=\sigma(x^c)$ such that $\xi$, $\eta$ and $X$  span a 3-dimensional Lie
algebra of KV's in a  spacetime ($M,\hat g$) where $\hat
g=e^{-2\sigma} g$; i.e.:  ($M,\hat g$) is a Bianchi type $III$ spacetime. One
can now adapt coordinates to  $\xi$, $\eta$ and $X$  in ($M,\hat g$) as
follows:
\begin{equation}
\xi={\partial \over \partial x^1}\qquad \eta=A{\partial \over \partial x^3} +
B{\partial \over \partial x^1}+ C{\partial \over \partial x^2} \qquad
X={\partial \over \partial x^3} 
\label{21} 
\end{equation}
where $A$, $B$ and $C$ are functions of $x^{\alpha}$, $\alpha=1,2,3$. The
commutation relations (\ref{20}) imply: $A=A(x^2)$, $B=x^1+B_0(x^2)$ and
$C=C(x^2)$; and one can then always carry out a change of coordinates
$(x^{\alpha})\rightarrow (x^{\alpha '})$ so as to write $\xi$, $\eta$ and $X$
as:
\begin{equation}
\xi={\partial \over \partial x^{1'}}\qquad \eta=
 x^{1'}{\partial \over \partial x^{1'}}+ {\partial \over \partial x^{2'}}
\qquad X={\partial \over \partial x^{3'}} 
\label{22} 
\end{equation}
 dropping now the primes and choosing a new coordinate $x^4$; it follows that
the metric $\hat g$ can be written as
\begin{equation}
 \hat g_{ab}= \left(\begin{array}{cccc}
e^{-2x^2} a_{11} & e^{-x^2} a_{12} & e^{-x^2} a_{13} & 0 \\
e^{-x^2} a_{12} &  a_{22} & a_{23} & 0 \\
e^{-x^2} a_{13} &  a_{23} & a_{33} & 0 \\
0 & 0  &  0  & \epsilon
\end{array}
\right) 
\label{23}
\end{equation}
where $\epsilon = \pm 1$ and $a_{\alpha \beta}=a_{\alpha \beta}(x^4)$. It is
now immediate to find out the general form for the metric $g$; for from
$g=e^{2\sigma}\hat g$ along with $\L_Xg=2g$ it readily follows that $\sigma =
x^3 + \sigma_0(x^4)$ (since $\L_{\xi}g=\L_{\eta}g=0$); redefining now the
coordinate $x^4$ one has
\begin{equation}
 g_{ab}= e^{2x^3}\left(\begin{array}{cccc}
e^{-2x^2} A_{11} & e^{-x^2} A_{12} & e^{-x^2} A_{13} & 0 \\
e^{-x^2} A_{12} &  A_{22} & A_{23} & 0 \\
e^{-x^2} A_{13} &  A_{23} & A_{33} & 0 \\
0 & 0  &  0  & \epsilon
\end{array}
\right) 
\label{24}
\end{equation}
where again $A_{\alpha\beta}=A_{\alpha\beta}(x^4)$ and $\epsilon=\pm 1$. The
case $\epsilon=+1$ corresponds to the 3-dimensional homothetic orbits being
timelike, whereas $\epsilon=-1$ corresponds to spacelike homothetic orbits.

\section{Examples}
The aim of this section is to provide a sample of physically significant
spacetimes admitting an $H_3$ on non-null orbits as maximal group of
similarity.

\subsection{Perfect Fluid Spacetimes}
A perfect fluid spacetime ($M,g$) satisfies the EFE's for an energy-momentum
tensor of the form:
\begin{equation}
T_{ab}= (\mu+p)u_au_b + pg_{ab}
\label{24b}
\end{equation}
where $u^a$ is the velocity flow of the fluid ($u^au_a=-1$), $\mu$ is a
positive function representing the energy density as measured by  an observer
comoving  with the fluid, and $p$ represents the pressure, usually satisfying
an equation of state of the form $p=p(\mu)$ (barotropic equation of state; the
fluid is then  said to be isentropic; i.e.: zero density of entropy
production). If a KV ${\cal X}$ exists in the spacetime, one has \cite{Kramer}:
\begin{equation}
\L_{{\cal X}}u_a=\L_{{\cal X}}\mu=\L_{{\cal X}}p=0
\label{25}
\end{equation}
and the existence of a proper HVF $X$ implies in turn
\cite{Wainwright,Eardley}:
\begin{equation}
\L_Xu_a=u_a\qquad \L_ X\mu=-2\mu \qquad p=( \gamma -1)\mu 
\label{26}
\end{equation}
where $ \gamma \in [1,2]$ in order to comply with the energy conditions
\cite{Kramer}. Henceforth, all of the examples we shall present will
correspond to the case of spacelike Killing orbits diffeomorphic to ${\cal
R}^2$ and the choice (\ref{11}) for the proper HVF. Thus, in the adapted
coordinate system set up in section 3.1 (see for example (\ref{6})) we shall
have:
\begin{equation}
\mu=e^{-2t}\hat\mu(z) \qquad u_t=-e^tf^{-1} \cosh \alpha(z) \qquad
u_z=e^tf^{-1} \sinh \alpha(z)
 \label{27}
\end{equation}
where $\alpha(z)$ is a function to be determined via the field equations. 
The components of the 4-velocity on the Killing orbits are zero ($u_x=u_y=0$)
also as a consequence of the field equations. As for the remaining cases
(timelike Killing orbits, Killing orbits diffeomorphic to a cylinder and/or the
ansatz (\ref{12}) for the proper HVF $X$), see remarks in the previous section
concerning this. Nevertheless, no correspondence will exist -in general-
between solutions obtained in all those various cases and those we will
present here (the only similarity being the general form of the metrics
under the changes of coordinates suggested in section 3, as it was
already pointed out there).

Going back to the current case, the generic form of the spacetime metric will
be that given in (\ref{6}) with the functions $\Psi$, $s$, $b$ and $P$ that
appear in (\ref{13})-(\ref{19}), depending on the particular Bianchi type we
are interested in. Furthermore, the generic forms of the fluid 4-velocity and
of the energy density will be those given by (\ref{27}). From the expression
of the velocity of the fluid, it is easy to see that, in general, it is
non-geodesic, expanding and shearing; its vorticity being zero since $u^a$ is
orthogonal to the Killing orbits.

At this point it is convenient to split up  our study into two
cases:

\subsubsection{The fluid flow velocity is tangent to the homothetic
orbits.}

In this case, and since we chose the coordinates $t$, $x$ and $y$ adapted to
the homothetic orbits; it follows that  $u_z=0$, and therefore the fluid is
comoving (for this particular choice of coordinates). One then has
\begin{equation}
u_t=-e^tf^{-1}(z) \qquad\dot u_z=-{f'\over f}
 \label{28}
\end{equation}
\begin{equation}
\theta=e^{-t}f(k+3)
 \label{29}
\end{equation}
where a dash indicates differentiation with respect to $z$, $\theta$ stands
for the expansion of the fluid ($\theta\equiv {u^a}_{;a}$); $k$ is defined
as $sb=e^{kt}\hat s(z)\hat b(z)$, and the remaining components of the 4-velocity
$u_a$ and the acceleration $\dot u_a$ are zero.

From the contracted Bianchi identities it follows
\begin{equation}
\gamma = {2 \over k+3}
 \label{30}
\end{equation}

Now, from the classification of
${\cal H}_3$ into Bianchi types, we see that for families $I$ and $II$
one has  $\gamma= {2\over 3}$. Such  a value for $\gamma$
lies out of the interval permitted ($\gamma \in [1,2]$); nevertheless, it is
physically significant since matter becomes attractive for $\gamma> {2\over
3}$; therefore the value ${2\over 3}$ may be of interest in inflationary
models \cite{Barrow}. For family $III$ one has $k=-1$ and
hence $\gamma=1$; i.e.: $p=0$ (dust). Families $IV$ and $V$
correspond to $k=-2$, i.e.: $\gamma=2$; that is: $p=\mu$
(stiff matter). For family $VI$, $k=-(q+1)$  (with $q\not=0,1$) and thus
$\gamma=2/(2-q)$; which taking into account the permitted values of
$\gamma$, implies that $q\in (0,1)$. For family $VII$, $k=-q$ (with $q^2<4$)
and therefore $\gamma=2/(3-q)$.

Furthermore, it is possible to see from the field equations, that the families
$(III)$ and $(V)$  admit no solutions of this type with $\mu \not=0$.

The case when $P=0$ (i.e.: ${\cal G}_2$ contains two mutually orthogonal KV's)
and $u^a$ is tangent to the homothetic orbits, has been thoroughly studied by
Wainwright and collaborators in a series of papers
\cite{Hewitt88,Hewitt90,Hewitt91} dedicated to investigate the role of
self-similarity in Cosmology.
They interpret these self-similar models as asymptotic states (at late times)
of more general inhomogeneous cosmological models; since they are precisely
those corresponding to the equilibrium points of the EFE's, written as an
autonomous system, for orthogonally transitive $G_2$ Cosmologies.

From our remarks above, it follows that their solutions must be of the type
$VI$ (type $I$ is ruled out since $\gamma \not\in[1,2]$, types $III$ and $V$
cannot admit solutions with $\mu\not=0$; and type $VII$  together with $P=0$
implies the existence  of a further KV tangent to the Killing orbits $V_2$ and
the metric would then admit a non-transitive group $H_4$ of homotheties).

Eardley \cite{Eardley} studied the case of 3-dimensional spacelike homothetic
orbits.

\subsubsection{ The fluid flow is not tangent to the homothetic orbits (tilted
case).}

In this case  $u_z\not=0$ for our particular choice of coordinates, and
consequently the expressions for the acceleration $\dot u_a$ and the expansion
$\theta$ of the fluid become more complicated, as well as the field equations.
In orthogonal transitive abelian $G_2$ models, it is possible, though,
to perform a change of coordinates in the $t$, $z$ plane so as to bring the
4-velocity of the fluid to a comoving form, preserving the diagonal form of
the induced metric there \cite{Beem}, as a consequence, the field equations can
be written in a much simpler form. Since most of the solutions of these
characteristics appearing in the literature are given in those coordinates, we
found convenient to translate our results (\ref{6}) and (\ref{13}-\ref{19}) to
them. Obviously, the form of the proper HVF $X$ will change and the
coordinates will no longer be adapted to the homothetic orbits; thus, we next
give the equivalents of  (\ref{6}) and (\ref{13})-(\ref{19}) in the new
coordinates. Following \cite{Wain}; the metric can now be written as:
\begin{equation}
ds^2= -A^2 dt^2 + B^2 dz^2  + r\{ f(dx +wdy)^2 + f^{-1}dy^2 \}
\label{31}
\end{equation}
where $A$, $B$, $r$, $f$ and $w$ are functions of $t$ and $z$ and the two
(commuting) KV's are $\xi={\partial\over \partial x}$ and $\eta={\partial
\over \partial y}$ (same as before).
The 4-velocity of the fluid is now:
\begin{equation}
u=A^{-1}{\partial\over \partial t}\, ,\quad {\rm or} \quad{\rm
equivalently}\quad u_a=(-A,0,0,0)
 \label{32}
\end{equation}
we can now use the remaining coordinate freedom in the $t$, $z$ plane
($t\longrightarrow m(t)$, $z\longrightarrow n(z)$) to bring the  (non-null)
proper HVF $X$ satisfying ($I-VII$) to either of the following three forms:
\begin{eqnarray}
(i)&\,&X=\partial_t + X^x(x,y)\partial_x + X^y(x,y)\partial_y 
\label{33}\\
(ii)&\,&X=\partial_z + X^x(x,y)\partial_x + X^y(x,y)\partial_y 
\label{34}\\
(iii)&\,&X=\partial_t+\partial_z + X^x(x,y)\partial_x + X^y(x,y)\partial_y 
\label{35}
\end{eqnarray}
$X^x(x,y)$ and  $X^y(x,y)$ being linear functions of the
coordinates $x$ and $y$, to be determined for each particular algebraic
Bianchi type $I$ to $VII$. Notice that ($i$) corresponds to $u$ being tangent to
the homothetic orbits, and therefore it has been dealt with above. ($ii$)
corresponds  to the orbits of the homothety group being spacelike (and also
$u^aX_a=0$); and this is the case studied by Eardley \cite{Eardley} (in this
case, and since $X$ and $u$ are mutually orthogonal it follows $\gamma=2$;
i.e.: $p=\mu$ stiff matter). Finally, ($iii$) is precisely the case we are
currently interested in; namely $u$ not tangent to the homothetic orbits.

Specializing now the equation (\ref{1}) to the metric (\ref{31}) and the HVF
$X$ given by (\ref{35}), we get for the metric functions
\begin{equation}
A^2=e^{t+z}\hat A^2(t-z) \qquad B^2=e^{t+z}\hat B^2(t-z)
\label{35b}
\end{equation}
in all seven types; and:
\begin{eqnarray}
(I)&\, & r=e^{t+z}\hat r(t-z)\qquad  f=\hat f(t-z)\qquad  w=\hat w(t-z) \\ 
(II)&\, & r=e^{t+z}\hat r(t-z)\qquad  f=\hat f(t-z)\qquad  w=\hat w(t-z)-{t+z
\over 2} \\  
(III)&\, & r=e^{{t+z\over 2}}\hat r(t-z)\qquad  f=e^{-{t+z\over 2}}\hat
f(t-z)\qquad  w=e^{{t+z\over 2}}\hat w(t-z) \\ 
(IV)&\, & r=\hat r(t-z)\qquad  f=\hat f(t-z)\qquad  w=\hat w(t-z)-{t+z
\over 2} \\  
(V)&\, & r=\hat r(t-z)\qquad  f=\hat f(t-z)\qquad  w=\hat w(t-z) \\  
(VI)&\, & r=e^{{1-q\over 2}(t+z)}\hat r(t-z)\quad  f=e^{-{1-q\over
2}(t+z)}\hat f(t-z)\quad  w=e^{{1-q\over 2}(t+z)}\hat w(t-z) \\ 
(VII)&\, & r=e^{{2-q\over 2}(t+z)} \sqrt{{4-q^2 \over 4}}\hat r(t-z) 
\end{eqnarray}
$$f={\sqrt{\hat r^2+\hat b^2+\hat c^2} +\hat b\cos(\sqrt{{4-q^2\over 4}}(t+z))
+\hat c\sin(\sqrt{{4-q^2\over 4}}(t+z)) \over  \sqrt{{4-q^2 \over 4}}\hat
r(t-z)}$$
$$w={q\over 2}+ \sqrt{{4-q^2 \over 4}} {\hat b\sin(\sqrt{{4-q^2\over
4}}(t+z)) - \hat c\cos(\sqrt{{4-q^2\over 4}}(t+z)) \over \sqrt{\hat r^2+\hat
b^2+\hat c^2} +\hat b\cos(\sqrt{{4-q^2\over 4}}(t+z)) +\hat
c\sin(\sqrt{{4-q^2\over 4}}(t+z))}$$
where $\hat b$ and $\hat c$ are both functions of ($t-z$).

It is interesting to notice that all diagonal ($w=0$), perfect fluid
solutions of the form (\ref{31}) (i.e.: admitting an orthogonally transitive
abelian $G_2$ with flat spacelike orbits) and such that the functions
$A$, $B$, $r$ and $f$ are separable in the variables $t$ and $z$ are already
known \cite{Ruiz}.
Note that the only Bianchi types which can contain diagonal metrics such
that, for them, the maximal isometry group is the abelian $G_2$ generated by
$\xi$ and $\eta$, are the types $I$, $III$, $V$ and $VI$ (families $II$ and
$IV$ do not contain diagonal metrics, and the diagonal, type $VII$ case admits a
further KV). For them, the metric functions are all of the form
\begin{equation}
F=e^{a(t+z)}\phi(t-z) \quad ,\quad a=constant
\label{43}
\end{equation}
and it is immediate to prove that $F$ is separable in $t$ and $z$ if and only
if $\phi$ is of the form:
\begin{equation}
\phi=Ce^{k(t-z)} \quad ,\quad C,k=constants
\label{44}
\end{equation}

We next present a few solutions which have been obtained for the Bianchi types
$I$, $III$ and $V$ assuming $w=0$ (diagonal), but which are not separable in
the above sense.\hfill\break

\noindent Type $I$
\begin{equation}
ds^2={e^{t+z}\over
{f_o}^2\vert 1-e^{-2(t-z)}\vert^{c^2}}\{-e^{t-z}dt^2+e^{-(t-z)}dz^2\}
+e^{2t}dx^2+ e^{2z}dy^2
 \label{45}
\end{equation}
$$\mu={c^2{f_o}^2\over e^{2t}}\vert 1-e^{-2(t-z)}\vert^{c^2}\quad  ,\quad
p=\mu$$

\noindent Type $III$
\begin{equation}
ds^2=e^{t+z} k^2 \vert 1-e^{-(t-z)}\vert^{ \beta} \{-e^{t-z}dt^2+dz^2\}
+e^{t+z}dx^2+ dy^2
 \label{46}
\end{equation}
$$\mu={1-\beta \over 4 k^2 e^{2t} \vert 1-e^{-(t-z)}\vert^{ \beta} }\quad 
,\quad p=\mu$$
 For $\beta=0$ in the above solution, the fluid has
geodesic flow ($\dot u_a=0$) and the metric admits a further spacelike KV
which is not tangent to the Killing orbits $V_2$; the solution being therefore
a special type of spatially homogeneous Bianchi cosmological model. For
$\beta\not=0$ the fluid is non-geodesic and the metric admits no further KV's.
\hfill\break

\noindent Type $V$
\begin{equation}
ds^2={e^{t+z}\over {f_o}^2} \left\{-dt^2{c^2 \varphi^2\over 1-c^2 \varphi^2}
+{dz^2\over 1- c^2 \varphi^2}\right\} + \varphi^2dx^2+ dy^2
 \label{47}
\end{equation}
$$\mu={1- c^2 \varphi^2\over 2c\varphi^2}{{f_o}^2\over e^{t+z}}\quad , \quad
p=\mu$$
 where $\varphi$ is a function of $t-z$ given implicitly by:
\begin{equation}
c(t-z)=\ln \varphi -{c^2\over 2}\varphi^2
\label{48}
\end{equation}
Notice that all these solutions have a stiff matter equation of state
($p=\mu$) and therefore can be derived from vacuum solutions (also admitting an
abelian $G_2$) using a method proposed by Wainwright et al. \cite{Wainwright79}

\subsection{Vacuum Solutions}
From our previous developments (see (\ref{13}) and (\ref{17})) it is possible
to find all vacuum solutions corresponding to types $I$ and $V$ in our
classification. Solving the vacuum field equations for them we get,
respectively:\hfill\break

\noindent Type $I$
\begin{eqnarray}
ds^2 & = & e^{2t}\bigg\{-dt^2 +dz^2 + {(e^{2z}-(\alpha^2+c^2) e^{-2z})^2\over
\alpha^2 e^{-2z}+(e^z-ce^{-z})^2}dy^2 +\\
   & + & (\alpha^2
e^{-2z}+(e^z-ce^{-z})^2)\left( {2\alpha\over \alpha^2
e^{-2z}+(e^z-ce^{-z})^2} dy+dx\right)^2 \bigg\} \nonumber
\end{eqnarray}
where $\alpha$, $c$ and $\beta$ are constants.\hfill\break

\noindent Type $V$
\begin{equation}
ds^2=e^{2t}e^{\alpha z} \{ -dt^2+dz^2\} + dy^2 +dx^2
\end{equation}
where $\alpha$ is a constant. It is immediate to see that this is (locally)
Minkowski spacetime and it can be brought to the standard form by means  of the
following coordinate change:
\begin{equation}
\hat t={1\over 2}\left( {e^{(1+\alpha)(t+z)}\over 1+\alpha} +
{e^{(1-\alpha)(t-z)}\over 1-\alpha}\right) \quad , \quad 
\hat z={1\over 2}\left( {e^{(1+\alpha)(t+z)}\over 1+\alpha} -
{e^{(1-\alpha)(t-z)}\over 1-\alpha}\right)
\end{equation}

 \vskip1cm
\noindent{\Large \bf Acknowledgments}
\vskip.8cm
The authors would like to thank Drs J.M.M. Senovilla (Universitat de
Barcelona) and G.S. Hall (University of Aberdeen) for many helpful discussions.
 Financial support from DGICYT Research project PB 91-0335 is
also acknowledged.

 \end{document}